\documentclass[acmsmall,screen,nonacm]{acmart}

\AtBeginDocument{%
  }

\authorsaddresses{}

\usepackage{amsmath,amsfonts,amsthm}
\usepackage[linesnumbered,ruled,vlined]{algorithm2e}
\usepackage{algorithmic}
\usepackage{graphicx}
\usepackage{lipsum}
\usepackage{wrapfig}
\usepackage{float}
\usepackage{textcomp}
\usepackage{colortbl}
\usepackage{xcolor}
\usepackage{xspace}
\usepackage{array}
\usepackage{soul}
\usepackage{url}
\usepackage{framed}
\usepackage{tcolorbox}
\tcbuselibrary{breakable}
\usepackage{hyperref}
\usepackage{listings}
\usepackage{tabularx}
\usepackage{makecell}
\usepackage{booktabs}
\usepackage{calc}
\usepackage{fancyvrb}
\usepackage{adjustbox}
\usepackage{multirow}
\usepackage{enumitem}
\usepackage{multirow}
    \acmConference[ISSTA 2026]{The ACM International Conference on Software Testing and Analysis}{3-9 Oct, 2026}{Oakland, California, United States}

\begin{document}
    \title{Understanding and Bridging the Planner-Coder Gap: A Systematic Study on the Robustness of Multi-Agent Systems for Code Generation}

    \author{Zongyi Lyu}
    \orcid{0009-0001-1600-4378}
    \affiliation{
      \institution{The Hong Kong University of Science and Technology}
      \city{Hong Kong}
      \country{China}
    }
    \email{zlyuaj@connect.ust.hk}

    \author{Songqiang Chen}
    \orcid{0000-0002-1220-8728}
    \affiliation{
      \institution{The Hong Kong University of Science and Technology}
      \city{Hong Kong}
      \country{China}
    }
    \email{i9s.chen@connect.ust.hk}

    \author{Zhenlan Ji}
    \authornote{Corresponding authors.}
    \orcid{0000-0003-3167-0480}
    \affiliation{
      \institution{The Hong Kong University of Science and Technology}
      \city{Hong Kong}
      \country{China}
    }
    \email{zjiae@cse.ust.hk}

    \author{Liwen Wang}
    \orcid{0009-0001-7831-6983}
    \affiliation{
      \institution{The Hong Kong University of Science and Technology}
      \city{Hong Kong}
      \country{China}
    }
    \email{lwanged@cse.ust.hk}

    \author{Shuai Wang}
    \authornote{Corresponding authors.}
    \orcid{0000-0002-0866-0308}
    \affiliation{
      \institution{The Hong Kong University of Science and Technology}
      \city{Hong Kong}
      \country{China}
    }
    \email{shuaiw@cse.ust.hk}

    \author{Daoyuan Wu}
    \orcid{0000-0002-3752-0718}
    \affiliation{
      \institution{Lingnan University}
      \city{Hong Kong}
      \country{China}
    }
    \email{daoyuanwu@ln.edu.hk}

    \author{Wenxuan Wang}
    \orcid{0000-0002-9803-8204}
    \affiliation{
      \institution{Renmin University of China}
      \city{Beijing}
      \country{China}
    }
    \email{jwxwang@gmail.com}
    
    \author{Shing-Chi Cheung}
    \orcid{0000-0002-3508-7172}
    \affiliation{
      \institution{The Hong Kong University of Science and Technology}
      \city{Hong Kong}
      \country{China}
    }
    \email{scc@cse.ust.hk}

    \newcommand{\parh}[1]{\noindent\textbf{#1}}
    \newcommand{\sparh}[1]{\noindent\underline{#1}}
    \newcommand{\F}{Fig.}
    \newcommand{\E}{Eq.}
    \newcommand{\T}{Table}
    \renewcommand{\S}{Sec.}
    \newcommand{\A}{Alg.}

    \newif\ifshowcomments
    \showcommentstrue 

    \ifshowcomments
        \definecolor{myblue}{RGB}{68,114,196} 
    \definecolor{mybrown}{RGB}{192,0,0} 
    \definecolor{mygreen}{RGB}{84,130,53}
    \newcommand{\fixme}[1]{{\color{red}{#1}}}
    \newcommand{\tocheck}[1]{{\color{orange}{#1}}}
    \newcommand{\todel}[1]{{\color{red}{\st{#1}}}}
    \newcommand{\sw}[1]{{\footnotesize{\textcolor{blue}{[SW: {#1}]}}}\xspace}
    \newcommand{\jzl}[1]{{\footnotesize{\textcolor{purple}{[JZL: {#1}]}}}\xspace}
    \newcommand{\csq}[1]{{\footnotesize{\textcolor{blue}{[CSQ: {#1}]}}}\xspace}
    \newcommand{\csqupd}[1]{#1}
    \newcommand{\lzy}[1]{{\footnotesize{\textcolor{myblue}{[LZY: {#1}]}}}\xspace}

    \else
    \newcommand{\fixme}[1]{#1}
    \newcommand{\tocheck}[1]{#1}
    \newcommand{\todel}[1]{#1}
    \newcommand{\sw}[1]{}
    \newcommand{\jzl}[1]{}
    \fi

    \begin{abstract}
Multi-agent systems (MASs) have emerged as a promising paradigm for automated
code generation, demonstrating impressive performance on established benchmarks
by decomposing complex coding tasks across specialized agents with different
roles. Despite their prosperous development and adoption, the fundamental
mechanisms underlying their robustness remain poorly understood, raising critical
concerns for real-world deployment.
This paper conducts a systematic empirical study to uncover the internal
robustness flaws of MASs using an automated mutation-based methodology. By
designing a testing pipeline incorporating semantic-preserving mutation
operators and a novel fitness function, we assess mainstream MASs
across multiple datasets and LLMs. Our findings reveal substantial
robustness flaws: semantically equivalent inputs cause drastic performance
drops, with MASs failing to solve 7.9\%--83.3\% of problems they
initially resolved successfully after applying the semantic-preserving
mutations.

Through comprehensive failure analysis, we discover a fundamental cause
underlying these robustness issues: the \textit{planner-coder gap}, which
accounts for 75.3\% of failures. This gap arises from information loss
in the multi-stage transformation process where planning agents decompose
requirements into underspecified plans, and coding agents subsequently
misinterpret intricate logic during code generation.
Based on this formulated information transformation process, we propose
a \textit{repairing method} that mitigates information loss through
multi-prompt generation and introduces a monitor agent to bridge the
planner-coder gap. Evaluation shows that our repairing method
effectively enhances the robustness of MASs by solving 40.0\%--88.9\% of identified
failures. Re-execution of the testing process on the repaired MASs shows that the number of detected failures
decreases up to 85.7\%, demonstrating that repaired MASs exhibit superior robustness.
Our work uncovers critical robustness
flaws in MASs and provides effective mitigation strategies, contributing
essential insights for developing more reliable MASs for code generation.
\end{abstract}


    
    \maketitle

\section{Introduction}

The advent of large language models (LLMs) has fundamentally transformed
code generation methodologies, enabling the automated production of code
that fulfills user requirements expressed in natural
language~\cite{jiang2024survey,hou2024large}. Among these methodologies,
multi-agent systems (MASs) have emerged as a particularly promising paradigm.
By leveraging the same backend LLM, MASs decompose complex coding tasks into
manageable sub-tasks and distribute them among specialized agents with
distinct roles~\cite{dong2024self,hong2023metagpt,zhang2024pair,
islam2024mapcoder,islam2025codesim}. For instance, \textit{planning agents}
design the overall coding logic and algorithmic structure, while \textit{coding
agents} implement the actual code based on these plans. Through carefully
crafted prompts and communication protocols, these specialized agents
collaborate effectively, achieving impressive performance
across diverse coding challenges on established benchmark
datasets~\cite{chen2021evaluating,austin2021program,
li2022competition,yu2024codereval}.
 
Despite their demonstrated effectiveness on benchmarks, a critical limitation exists
in our understanding of MAS robustness for code generation. While MASs work
well on standard evaluation based on test cases, the fundamental mechanisms underlying their
failures remain poorly understood---particularly when faced with semantically
equivalent variations of input requirements. This ``black box'' problem poses
significant risks for real-world deployment: non-robust MASs may generate
inconsistent or incorrect code solutions, resulting in system failures and
increased debugging costs~\cite{mastropaolo2023robustness,zhong2024lcb}.
Without a systematic understanding of \textit{why} and \textit{how} MASs fail
under semantic-preserving input variations, it remains difficult to develop
principled solutions for enhancing their reliability.

To address this knowledge gap, we conduct a systematic empirical study to
uncover the internal robustness flaws of MASs using an automated
mutation-based methodology. Inspired by existing efforts on AI model
robustness testing~\cite{wang2023recode,mastropaolo2023robustness,
chen2021testing,iakusheva2023metamorphic,xie2022boosting,
rehman2023metamorphic}, we design four semantic-preserving mutation operators
that systematically transform input questions while preserving their intended
semantics. Our testing methodology is guided by a novel fitness function that
evaluates both the quality of generated code and the semantic similarity of
generated plans, providing fine-grained feedback for mutation selection. We
assess three mainstream MASs~\cite{dong2024self,hong2023metagpt,zhang2024pair}
across multiple datasets and backend LLMs. Our empirical findings reveal
substantial robustness flaws: semantically equivalent inputs cause drastic
performance drops, with MASs failing to solve 7.9\%--83.3\% of problems they
initially resolved successfully after applying semantic-preserving mutations.

Through comprehensive failure analysis of the testing results, we discover a
fundamental cause underlying these robustness issues: the
\textit{planner-coder gap}, which accounts for 75.3\% of all observed
failures. This gap arises from information loss in the multi-stage
transformation process inherent to MAS architectures. Specifically, planning
agents decompose user requirements into high-level plans that are often
logically sound but lack sufficient implementation details, while coding
agents subsequently misinterpret intricate logic or ambiguous expressions
during code generation. We systematically characterize this gap through five
distinct error patterns (EPs): \textit{EP-1: Gap in Core Concepts},
where coding agents misunderstand fundamental algorithmic concepts in plans;
\textit{EP-2: Gap in Edge Cases}, where boundary conditions specified in plans
are overlooked during implementation; \textit{EP-3: Gap in Complex Logic},
where multi-step reasoning or nested conditions are incorrectly
translated; \textit{EP-4: Gap in Relational Phrases}, where quantity or spatial relationships are misinterpreted; and \textit{EP-5:
Gap in Condition Judgments}, where conditional branches or guard clauses are
improperly implemented. These patterns reveal that enhancing MAS robustness
fundamentally requires bridging the planner-coder gap to mitigate information
loss across agents.

Based on our formulated understanding of the information transformation process,
we propose a repairing method to mitigate information loss and bridge the
planner-coder gap. Our approach incorporates two synergistic components. First,
\textit{multi-prompt generation} diversifies input question expressions through
semantic-preserving mutations, reducing the likelihood of misinterpretation.
Second, we introduce a specialized \textit{monitor agent} as an intermediary
between planning and coding agents. This monitor agent operates through two key
processes: (1) a \textit{plan interpretation process} that provides detailed
explanations to enhance the coding agent's comprehension, and (2) a \textit{code
validation process} that verifies alignment between generated code and the
interpreted plan. By explicitly addressing information loss at the critical
planner-coder interface, our monitor agent effectively bridges the communication
gap underlying most robustness failures.


We conduct extensive experiments to evaluate our repairing method across
three prominent MASs~\cite{dong2024self,hong2023metagpt,zhang2024pair}
with three backend LLMs~\cite{achiam2023gpt,zhu2024deepseek,
openai2023chatgpt} on four datasets~\cite{chen2021evaluating,austin2021program,
li2022competition,yu2024codereval}. Results demonstrate that our approach
successfully repairs 40.0\%--88.9\% of identified failures. When we re-execute
the testing process on repaired MASs, discovered failures decrease by up to
85.7\%, providing empirical evidence of substantially improved robustness.
Ablation studies confirm that both components are necessary, as they address
complementary aspects of the robustness challenge.
Overall, we summarize our contributions as follows:

\begin{itemize}[leftmargin=*,noitemsep,topsep=0pt]
    \item We conduct the first systematic empirical study on the robustness of
 MASs for code generation, focusing on their behavior under semantically
 equivalent input variations. Our automated mutation-based methodology
 incorporates semantic-preserving mutations and a novel fitness function to
 effectively assess mainstream MASs, revealing substantial robustness flaws
 where performance drops by 7.9\%--83.3\% under semantic-preserving
 transformations.

    \item We discover and characterize the \textit{planner-coder gap} as a
 fundamental cause of MAS robustness failures, accounting for 75.3\% of
 observed issues. Through systematic analysis, we identify five distinct
 error patterns that manifest this gap and formulate the multi-stage
 information transformation process in MAS code generation, providing
 theoretical grounding for understanding information loss across agent
 boundaries.

    \item Based on our empirical findings and theoretical formulation, we
 propose a repairing method that mitigates information loss through
 multi-prompt generation and monitor agent insertion. Comprehensive
 experiments demonstrate that our approach repairs 40.0\%--88.9\% of
 identified failures and reduces failure rates by up to 85.7\% in subsequent
 testing, validating its effectiveness in enhancing MAS robustness. 
\end{itemize}

    \section{Background}
\label{sec:background}


Code generation has been a central focus of software engineering research,
aiming to automate programming tasks and enhance developer productivity by
generating desired code from natural language
specifications~\cite{shin2021survey,
dehaerne2022code,cao2024javabench,du2024evaluating,mathews2024test,gao2024preference}.
MASs are artificial intelligence systems that leverage LLMs as their core
computational engines to simulate domain experts with specialized
capabilities~\cite{qian2023communicative, hong2023metagpt,zhang2024pair}. With advances in prompt engineering~\cite{liu2023pre},
developing MASs to enhance code generation has emerged as a prominent
research direction~\cite{he2024llm,tran2025multi}.

\begin{figure}[!tbp]
    \centering
    \includegraphics[width=0.55\textwidth]{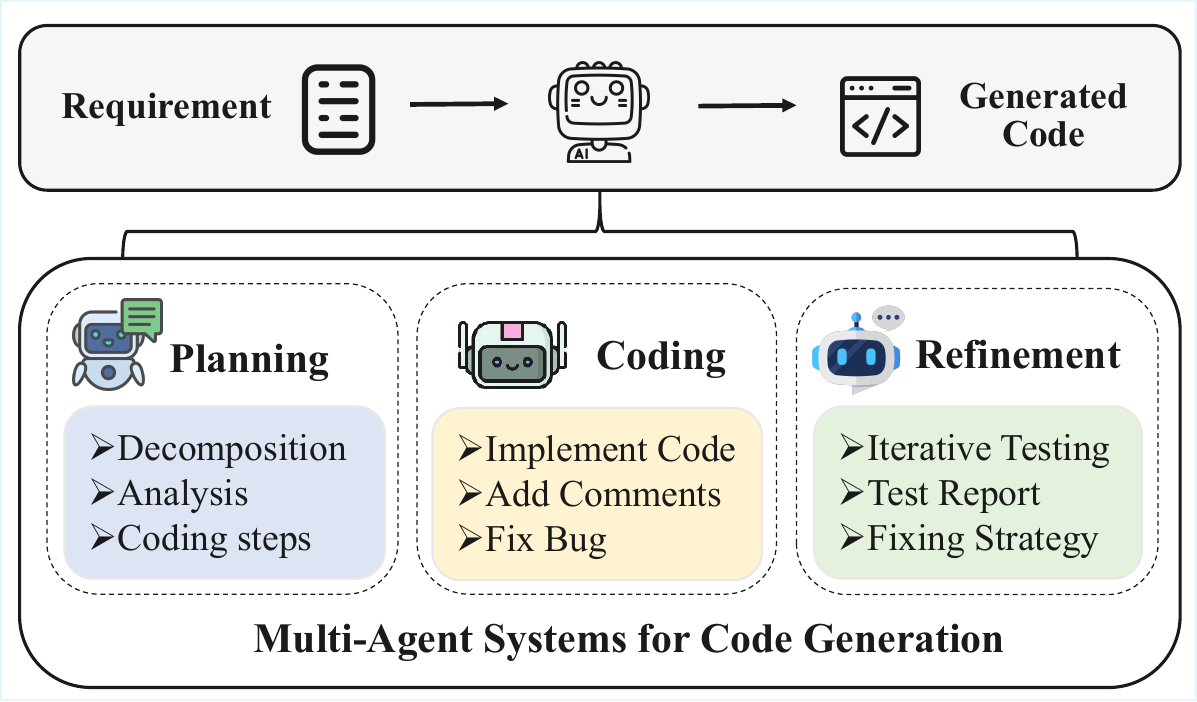}
    \vspace{-8pt}
    \caption{Pipeline of MASs for code generation. }
    \label{fig:mas4code}
    \vspace{-16.2pt}
\end{figure}

A prevalent multi-agent approach to code generation employs role specialization
and iterative feedback mechanisms to provide code snippets that satisfy the user input~\cite{liu2024large}. As shown in \F~\ref{fig:mas4code}, the
typical architecture for code generation MASs consists of three primary parts:
planning, coding, and feedback-driven refinement, which are typically implemented through planner, coder, and tester~\cite{he2024llm}. 
\F~\ref{fig:mas4code} depicts a typical workflow of a code generation MAS.
Upon receiving a natural language requirement, the system routes the input to
the \textit{planning agent} (planner) to generate a clear and structured coding plan. Note that the planning agent does not simply
paraphrase the user input. Instead, it serves as an important coding step that decomposes the coding problem
into a set of requirements and designs several key logical steps to solve the problem,
which determines the algorithm and the logic for code
implementation~\cite{zhang2024pair,dong2024self}.
The \textit{coding agent} (coder) then implements the solution following the
provided plan. Subsequently, the \textit{testing agent} (tester) automatically generates test
cases and evaluates the implementation, producing a test report. If the code
fails, the report is returned to the coder for iterative refinement until the
solution passes all tests or the computational budget is exhausted.


This specialized multi-agent workflow represents a fundamental paradigm shift from single-LLM code generation approaches. Unlike the end-to-end paradigm where a unified model directly transforms natural language into 
executable code~\cite{wang2021codet5,fried2022incoder}, MASs decompose the complex task into specialized subtasks handled by domain-specific agents. 
This paradigm demonstrates remarkable advantages in handling complex coding tasks. Nevertheless, our analysis in Sec.~\ref{sec:implication} reveals critical
robustness issues concerning the interaction between planner and coder. 








\section{Motivation}
\label{sec:motivation}

MASs~\cite{hong2023metagpt,dong2024self,zhang2024pair} have emerged as
a transformative technology with immense commercial potential. IT
companies have invested significant effort in the development of MASs, with Google Cloud
announcing comprehensive enhancements to Vertex AI that include the Agent
Development Kit (ADK) and Agent Engine for building and deploying
enterprise-grade MASs at scale~\cite{cloud2025mas}. OpenAI has also
launched new APIs and tools specifically designed for agentic applications,
including the Responses API and the Agents SDK for orchestrating
multi-agent workflows, demonstrating growing commercial interest in
MASs~\cite{openai2025mas}. 

However, assessing the reliability and robustness of these complex systems 
remains a significant research gap. Unlike single agents and individual models, MASs involve intricate
inter-agent communications, dynamic role assignments, and collaborative
decision-making processes that make their behavior difficult to predict and
evaluate systematically~\cite{he2024llm}. This gap in assessment capabilities becomes
particularly concerning given the high-stakes commercial
applications these systems are designed to serve~\cite{he2024llm}.
Inspired by previous work that employs
testing to assess the robustness of AI
systems~\cite{li2023cctest,mastropaolo2023robustness},
we assess the robustness of MASs for code generation
based on semantically equivalent input mutations.
In particular, we presume that a robust MAS should maintain consistency
when addressing semantically equivalent input and generating correct programs
regardless of how developers articulate their requirements. 
This idea stems from the observation that developers' requests typically
vary in vocabulary and phrasing due to the variability of natural
language.

To illustrate this idea, consider a simple example where a developer might
request: (1) ``return the n-th Fibonacci number'', (2) ``give back the n-th number
in the Fibonacci sequence.'' Although these requests use different phrasing,
their semantic meaning is identical. A robust MAS should generate functionally
correct solutions for both expressions. However, we find that even the
state-of-the-art MASs suffer from
inconsistencies in handling such semantically equivalent expressions. For example, PairCoder~\cite{zhang2024pair}
generates correct code for the first request, while failing on the second due to incorrect edge case handling.

    \section{Fuzzing MAS with Semantically Equivalent Requirements}
\label{sec:fuzzing}

To fill the gap in understanding of MASs' robustness in code generation as
discussed in Sec.~\ref{sec:motivation}, we design a fuzz testing framework
specifically adapted to MAS architectures. Our approach systematically generates
semantically equivalent variations of programming requirements to expose
robustness flaws arising from inter-agent communication---a vulnerability
dimension not addressed by traditional testing approaches. A key innovation is
our \textbf{dual-component fitness function}, which evaluates mutations along
two dimensions: (1) final code quality degradation and (2) semantic drift in
intermediate plans. This MAS-specific design enables us to systematically uncover
the planner-coder gap, a fundamental structural flaw that represents the
dominant failure mode in contemporary MASs (see \S~\ref{sec:implication}).

\subsection{Fuzzing Methodology}

We first present four semantic-preserving mutation operators that generate
alternative expressions of selected questions. Then, we introduce a novel fitness
function that calculates rewards based on the differences in both output code
and generated plans between the original and mutated questions, to provide
comprehensive guidance for subsequent fuzzing iterations.

\parh{Mutation Operators.} 
\label{sec:mutation_operators}
Our mutation approach focuses on generating diverse expressions while preserving
the semantic meaning of input questions. To achieve this, we design four sentence-level
mutation operators that modify the expression styles of natural language requirements without altering the original
semantic content.



\begin{table}
    \centering
    \caption{Semantic-preserving mutation operators and descriptions.}
    \begin{adjustbox}{width=1\linewidth}
    \begin{tabular}{l|l}
    \toprule
    \textbf{Operator} & \textbf{Description} \\
    \midrule
    Rephrase & Instruct an LLM to rewrite a sentence using other words while maintaining the overall meaning. \\
    \midrule
    Insert & Prompt an LLM to append one additional sentence at the end of the description based on the semantic content. \\
    \midrule
    Expand & Use an LLM to expand one sentence into two by distributing its semantic content.  \\
    \midrule
    Condense & Apply an LLM to condense two consecutive sentences into one sentence using appropriate conjunctions.\\
    \bottomrule
    \end{tabular}
    \end{adjustbox}
    \label{tab:operators}
    \vspace{-10pt}
\end{table}

Table~\ref{tab:operators} presents these four mutation operators: \textit{Rephrase},
\textit{Insert}, \textit{Expand}, and \textit{Condense}. We implement these
operators using GPT-4o~\cite{achiam2023gpt}, a state-of-the-art LLM, through
carefully designed prompts.
This design effectively addresses two key challenges from previous
work~\cite{wang2023recode, mastropaolo2023robustness}: (1) it minimizes semantic deviations that may occur during rule-based mutation, and (2) it produces natural and
fluent expressions in the mutated questions.
Moreover, as shown in prior work~\cite{yu2024llm}, LLM-generated mutations
effectively simulate diverse human expressions. Our operators cover prevalent
semantic-preserving transformations~\cite{yu2024llm}, providing comprehensive
coverage of semantic variations.

To confirm the reliability of our mutation operators in terms of semantic
preservation, we conduct manual verification on 500 randomly sampled mutants, confirming that 99.2\% of the mutants preserve semantics. 
Additionally, we measure the semantic similarity of all mutants to the original
questions using Sentence-Transformers~\cite{reimers2019sentence}, whose results
show 98.4\% average similarity.

\parh{Fitness Function.} To determine whether a mutated question should be retained in the seed pool
for further mutation, we propose a novel fitness function that calculates
rewards for MAS outputs on mutated questions. Unlike previous coverage-driven fuzzing
approaches~\cite{zalewski2023afl,bohme2016coverage,ognawala2018improving,veggalam2016ifuzzer} on software, 
our fitness function considers the
multi-agent nature of MASs by comparing the output quality between the seed
questions and their mutated variants across two distinct aspects: code
and plan.


\sparh{Code Reward.}~To mitigate the occasional failure 
due to randomness rather than underlying robustness issues, 
we execute MASs for $n$ times for each question.
We collect the generated code and calculate the rate of
completions passing all test cases provided in the dataset.
Based on the insight that an increasing number of failure trials on the mutated
questions indicate degraded MAS
performance~\cite{wang2023codet5+,chen2021evaluating}, we calculate the code
reward by measuring the difference between the pass rates of code generated
from original versus mutated questions:
\begin{equation}
 \mathcal{R}_{C} = \frac{1}{n} \left(\sum_{i=1}^{n} c_i - \sum_{i=1}^{n} \hat{c}_i\right)
\end{equation}

\noindent where $n$ represents the number of attempts, $c_i$ and $\hat{c}_i$
indicate whether the $i$-th generation passes the test (1 for pass, 0 for
fail). Positive rewards indicate that the mutated question reduces MAS
performance, making it valuable for identifying robustness issues
in subsequent fuzzing processes.

\sparh{Plan Reward.}~Apart from the generated code, we also compare the semantic
of plans generated by the planning agent for the original and mutated
questions. Significant changes in the generated plans often indicate potential
robustness issues in the planning phase, which substantially influence 
the coding agent's ability and the final code quality. We apply
Sentence-Transformers~\cite{reimers2019sentence} to obtain the embedding and calculate
semantic similarity between corresponding plans. The plan reward is computed by
subtracting the normalized similarity from 1:
\begin{equation}
 \mathcal{R}_{P} = \frac{1}{n} \sum_{i=1}^{n} \left(1 - \frac{\hat{p}_i \cdot p_i}{||\hat{p}_i|| \cdot ||p_i||}\right)
\end{equation}

\noindent where $p_i$ and $\hat{p}_i$ represent the vector embeddings of plans
generated from the seed question and the mutated question, respectively, for the
$i$-th generation.
 
The final fitness function is calculated as the sum of the code reward and the
plan reward. 
This combination enables measuring robustness issues in both the code quality degradation and semantic drift in the planning stage, which is the key to comprehensively understanding MAS robustness.



\begin{figure}[t]
    \centering
\begin{algorithm}[H]
    \footnotesize
    \caption{Fuzzing Pipeline}
    \label{algo:fuzzing}
    \KwIn{Multi-Agent System $M$, Seed Pool $\mathcal{S}$, Fitness Function $\mathcal{F}$, Budget $\mathcal{B}$, Attempts $n$}

    \KwOut{Failure Question Set $\mathcal{R}$}

    $\mathcal{R} \gets \emptyset$ \;

    \While{$ConsumedQueries < \mathcal{B}$ and $\mathcal{S} \neq \emptyset$}{
        $seed \gets \text{MCTS-exploring}(\mathcal{S})$ \;
        $mutated\_seed \gets \text{Mutate}(seed);$ \tcp{Mutate the seed question with randomly selected operator}

        $failures \gets 0$\;
        \For{$i = 1$ \KwTo $n$}{
            $result \gets \text{Execute}(M, mutated\_seed) ;$  \tcp{Execute the MAS with the mutated seed}
            \If{$result = \text{False}$}{  
                $failures \gets failures + 1$\;
            }
        }

        \If{$failures = n$}{  
            $\mathcal{R} \gets \mathcal{R} \cup \{mutated\_seed\};$ \tcp{Add the mutated seed to the failure set}
            $original\_seed \gets  \text{GetParent}(seed);$ \tcp{Get the original seed question} 
            $\mathcal{S} \gets \mathcal{S} \setminus \{original\_seed\};$ \tcp{Stop fuzzing on this branch}
        }
        \Else{
            $reward \gets \mathcal{F}(result);$ \tcp{Calculate reward using fitness function}
            \If{$reward > 0$}{ 
                $\mathcal{S} \gets \mathcal{S} \cup \{mutated\_seed\};$ \tcp{Add the mutated seed to the seed pool} 
            }
        }
    }
    \KwRet $\mathcal{R}$  
\end{algorithm}
\vspace{-13pt}
\end{figure}

\parh{Fuzzing Pipeline.} Following prior fuzzing
methodologies~\cite{fioraldi2022libafl,zalewski2023afl}, our approach
encompasses four major components: seed pool initialization, seed
selection, mutation, and execution.
The workflow of our fuzzing process is presented in \A~\ref{algo:fuzzing}.
The algorithm takes the target MAS $\mathcal{M}$, seed pool $\mathcal{S},$
fitness function $\mathcal{F}$, fuzzing budget $\mathcal{B}$ and total attempts $n$ as inputs, and outputs failure question set
$\mathcal{R}$.
Specifically, we first create an initial seed pool from questions in
established datasets that the target MAS successfully resolves. For
each fuzzing iteration, following recent advances in fuzzing
methodologies, we employ
MCTS-exploring~\cite{yu2024llm} as our seed selection mechanism (line
3), which effectively identifies promising seeds while avoiding local
optima. Once a seed is selected, we randomly choose one of our four
semantic-preserving mutation operators to generate a semantically
equivalent version of the question (line 4). Subsequently, the
mutated question is submitted to the MAS, which executes the code
generation process $n$ times (where $n = 10$ in our implementation) and
records the failure numbers (lines 6--9). If the
MAS fails to correctly solve the mutated question in all trials, we
identify this as an unsolvable variant and halt further fuzzing on that
branch (lines 11--13). If the MAS successfully handles the mutated
question, we calculate the reward using our fitness function to
determine whether to retain the mutation for further exploration (lines
15--17). We continue the above process until we reach predefined
resource constraints or exhaust valid seeds for mutation.

\subsection{Experiment Setup}
\label{sec:fuzzing_setup}

\parh{MASs.} We evaluate three widely-adopted code generation MASs:
Self-Collaboration Code Generation~\cite{dong2024self} (\textit{abbr.} SCCG),
MetaGPT~\cite{hong2023metagpt}, and PairCoder~\cite{zhang2024pair}. All
systems incorporate agents responsible for planning and coding. MetaGPT
represents one of the earliest MAS frameworks, applying five agents to
collaboratively develop software based on user instructions. SCCG proposes the workflow of planning, coding and
testing, building a popular framework for many subsequent works~\cite{zhang2024pair,islam2024mapcoder}. PairCoder is the state-of-the-art MAS that includes plan selection.

\parh{Backend LLMs.}
MASs employ the same backend LLM for all constituent agents. We use three
different LLMs as backend models for the aforementioned MASs:
GPT-3.5-Turbo~\cite{openai2023chatgpt} (\textit{abbr.} GPT-3.5),
GPT-4o~\cite{achiam2023gpt}, and Deepseek-Coder-V2~\cite{zhu2024deepseek}
(\textit{abbr.} Deepseek). All these LLMs are widely adopted and demonstrate
strong capabilities in code
generation~\cite{joel2024survey,fakhoury2024exploring,jiang2024survey}.
We apply the same parameters for the backend LLMs as the original MAS
papers~\cite{dong2024self,hong2023metagpt,zhang2024pair}.

\parh{Datasets.}~We employ four code generation datasets: HumanEval 
ET~\cite{chen2021evaluating} (\textit{abbr.} HumanEval), MBPP ET~\cite{austin2021program} (\textit{abbr.} MBPP),
CodeContest~\cite{li2022competition} and CoderEval~\cite{yu2024codereval}, which
are widely used for benchmarking MASs for code
generation~\cite{dong2024self,lin2024soen}. HumanEval and MBPP~\cite{dong2025codescore} 
provide well-constructed coding problems, 
serving as standard benchmarks. 
CodeContest~\cite{li2022competition} represents competitive programming
scenarios, offering more challenging problems. Notably, CoderEval~\cite{yu2024codereval} is latest dataset that mirrors real-world development with data sourced from actual GitHub repositories, thereby offering a modern and realistic assessment of MASs.
These datasets, ranging from classical benchmarks to cutting-edge real-world scenarios, are chosen for evaluation across various coding contexts.

\parh{Dataset Splitting.}~It is important to note that we apply 50\% of
the data in each dataset for testing, and the rest for repairing (see
\S~\ref{sec:repairing}). In other words, this strictly prohibits using the same
data for both fuzzing and repairing in order to avoid data leakage. We use the
sanitized version of MBPP, the test set of CodeContest, and
the standalone-level of CoderEval to save resources.

\parh{Evaluation Metric and Fuzzing Parameters.}~We adopt
Pass@\textit{k}~\cite{chen2021evaluating} as our primary metric to
comprehensively assess MAS robustness. Following previous
works~\cite{chen2021evaluating,li2022competition}, we
set $k = 10$ in our evaluation. Compared to Pass@1, Pass@10 mitigates the
influence of occasional failures due to randomness and can reflect more nuanced performance
differences among MASs~\cite{austin2021program}. To balance comprehensiveness and resource consumption, 
we set the query budget of our fuzzing to 10,000. 
For the MCTS-Explore algorithm, we apply identical parameter settings 
as specified in~\cite{yu2024llm}. To prevent local optima, 
after a question is selected 15 times, we terminate fuzzing on the branch.



\subsection{Fuzzing Results}
\label{sec:fuzzing_results}

\begin{table*}[!tb]
    \setlength{\tabcolsep}{1.5pt}
    \caption{Performance comparison on the original performance and performance after fuzzing.}
    \centering
    \resizebox{1.0\linewidth}{!}{
        \begin{tabular}{c|c|ccc|ccc|ccc|ccc}
            \toprule
            \multirow{2}{*}{MAS}                & \multirow{2}{*}{Backend LLM} & \multicolumn{3}{c|}{{HumanEval }} & \multicolumn{3}{c|}{{MBPP }} & \multicolumn{3}{c|}{{CodeContest}}  & \multicolumn{3}{c}{{CoderEval}}                                                                                        \\
            \cmidrule(lr){3-5} \cmidrule(lr){6-8} \cmidrule(lr){9-11} \cmidrule(lr){12-14}
                                                &                              & Fuzzing                             & Original                       & Drop                              & Fuzzing & Original & Drop                & Fuzzing & Original & Drop                & Fuzzing & Original & Drop   \\
            \midrule
            \multirow{3}{*}{SCCG}               & GPT-3.5                      & 0.4815                              & 0.7160                         & $\downarrow$ 32.6\%               & 0.4590  & 0.5995   & $\downarrow$ 23.4\% & 0.0364  & 0.1091   & $\downarrow$ 66.7\% & 0.2540  &  0.4286  & $\downarrow$ 40.7\%\\
                                                & GPT-4o                       & 0.6975                              & 0.8395                         & $\downarrow$ 16.9\%               & 0.5761  & 0.6909   & $\downarrow$ 16.6\% & 0.2121  & 0.3394   & $\downarrow$ 37.5\% & 0.3016  &  0.4921  & $\downarrow$ 38.7\%\\
                                                & Deepseek                     & 0.5617                              & 0.7345                         & $\downarrow$ 23.5\%               & 0.4964  & 0.6440   & $\downarrow$ 22.9\% & 0.0242  & 0.0970   & $\downarrow$ 75.0\% & 0.1429  &  0.3810  & $\downarrow$ 62.5\%\\

            \midrule

            \multirow{3}{*}{MetaGPT}            & GPT-3.5                      & 0.4085                              & 0.7073                         & $\downarrow$ 42.2\%               & 0.5761  & 0.6651   & $\downarrow$ 13.4\% & 0.0242  & 0.0788   & $\downarrow$ 69.2\% & 0.3016  &  0.4127  & $\downarrow$ 26.9\%\\
                                                & GPT-4o                       & 0.6707                              & 0.8170                         & $\downarrow$ 17.9\%               & 0.6089  & 0.6979   & $\downarrow$ 12.8\% & 0.1879  & 0.3091   & $\downarrow$ 39.2\% & 0.3333 &  0.4603  & $\downarrow$ 27.6\%\\
                                                & Deepseek                     & 0.5915                              & 0.7378                         & $\downarrow$ 19.8\%               & 0.5972  & 0.6812   & $\downarrow$ 12.9\% & 0.0242  & 0.1515   & $\downarrow$ 83.3\% & 0.2540  &  0.3968  & $\downarrow$ 36.0\%\\

            \midrule
            \multirow{3}{*}{PairCoder}          & GPT-3.5                      & 0.6829                              & 0.7744                         & $\downarrow$ 11.8\%               & 0.6721  & 0.7307   & $\downarrow$ 8.0\%  & 0.1030  & 0.2364   & $\downarrow$ 43.6\% & 0.0952  &  0.4127  & $\downarrow$ 76.0\%\\
                                                & GPT-4o                       & 0.7622                              & 0.8475                         & $\downarrow$ 10.1\%               & 0.6791  & 0.7377   & $\downarrow$ 7.9\%  & 0.4424  & 0.5273   & $\downarrow$ 16.1\% & 0.1587  &  0.4762  & $\downarrow$ 66.7\%\\
                                                & Deepseek                     & 0.7683                              & 0.8536                         & $\downarrow$ 10.0\%               & 0.6417  & 0.7025   & $\downarrow$ 8.7\%  & 0.1576  & 0.2788   & $\downarrow$ 43.5\% & 0.2698  &  0.4603 & $\downarrow$ 41.4\%\\
            \bottomrule
        \end{tabular}
    }
    \label{tab:fuzzing_result}
    \vspace{-13pt}
\end{table*}

\T~\ref{tab:fuzzing_result} presents the performance differences of MASs
before and after fuzzing. The \textit{Drop} represents the decrease rate of
pass@10 after fuzzing. Our results reveal that all MASs with
different backend LLMs experience performance degradation, with pass@10
decreasing by 7.9\%-- 83.3\%. Specifically, SCCG (GPT-3.5) on CodeContest
demonstrates the highest accuracy drop, while PairCoder (GPT-4o) on MBPP
exhibits the best robustness. These findings demonstrate that current MASs
suffer from robustness issues when confronted with semantically equivalent
questions, despite achieving promising results on original datasets.

Our fuzzing results reveal that different MASs exhibit varying degrees of robustness issues. 
PairCoder consistently demonstrates the best robustness across most datasets, 
while SCCG and MetaGPT suffer from more severe performance degradation. 
For instance, when using GPT-3.5 as the backend LLM on HumanEval , 
PairCoder's performance drops by only 11.8\%, whereas SCCG and MetaGPT 
experience over 30\% decline. The superior robustness of PairCoder can be 
attributed to its navigator agent that generates and selects optimal plans 
from multiple candidates.
However, PairCoder's advantage diminishes on CoderEval, where it 
suffers the largest performance drop due to its heavy reliance on public 
test cases for iterative code refinement, which CoderEval does not provide.
Meanwhile, MetaGPT shows better robustness on MBPP  
and CoderEval, which contain shorter questions with minimal
explanations, aligning with MetaGPT's design philosophy of processing brief user requirements. 
These findings suggest that plan selection mechanisms and dataset characteristics 
may influence MAS robustness.

Moreover, we notice that MASs exhibit more severe robustness issues when dealing with 
longer and more complex questions. Among the four datasets, CodeContest and 
CoderEval exhibit substantially larger performance drops than HumanEval 
and MBPP. This pattern can be explained by their distinct characteristics: 
CodeContest contains competition-level problems with lengthy descriptions 
and strict input format constraints, while CoderEval involves intricate 
problem settings that mirror real-world scenarios. Our findings reveal 
that MASs are particularly sensitive to these questions, where even 
minor perturbations can trigger failures. This heightened sensitivity 
raises concerns about the reliability of MASs in practical applications.





We further observe that the capability of the backend LLM affects MAS
robustness, with powerful LLMs demonstrating superior robustness. Across
all datasets, MASs powered by GPT-3.5 experience more significant performance
degradation than those using GPT-4o. We attribute this difference to the limited
semantic understanding of weaker LLMs, which makes them less capable of
accurately interpreting diverse expressions and thus more vulnerable to
robustness issues.



\begin{tcolorbox}[ size=small , enlarge top by=-4pt, enlarge bottom by=-3pt] 
 \textbf{Our Findings:}~Existing MASs generally suffer from robustness
 issues, manifesting a substantial performance drop when
 processing semantically equivalent questions. This problem is
 further exacerbated by longer and more complex questions, but can be
 mitigated by using more powerful backend LLMs. Furthermore, our
 results reveal that plan selection and refinement mechanisms may
 enhance MAS robustness, indicating a potential direction to improve
 MAS design.
\end{tcolorbox}
    \section{Implications from Failure Cases}
\label{sec:implication}

The substantial robustness flaws revealed in 
\S~\ref{sec:fuzzing}---where semantically equivalent mutations cause performance drops of 7.9\%--83.3\%---raise a fundamental question: \textit{what underlying mechanisms cause MASs to fail on inputs they should handle correctly?}
To answer this question, we conduct a systematic failure analysis on over 700 uncovered failures.
Specifically, we randomly sample 20\% of failures
and manually examine both the generated code and internal outputs (e.g., plans) to
investigate the root causes.
To mitigate subjectivity, two software developers first build a
failure taxonomy through a pilot study on 5\% of cases, then independently
categorize the remaining cases. The final Cohen's Kappa
score~\cite{cohen1960coefficient} is 0.88, indicating high agreement.


\begin{wraptable}{r}{0.4\textwidth}
    \centering
    \vspace{-5pt}
    \caption{Distribution of reasons for failures.}
    \scalebox{0.8}{
    \begin{tabular}{l|cc}
        \toprule
        \textbf{Category} & \textbf{\#Failure} & \textbf{Percentage} \\
        \midrule
        Planner-Coder Gap & 113                     & 75.3\%              \\
        Plan Logic Errors & 23                      & 15.3\%              \\
        Invalid           & 14                      & 9.3\%               \\
        \bottomrule
    \end{tabular}
    }
    \label{tab:failure_distribution}
    \vspace{-5pt} 
\end{wraptable}

Our systematic analysis yields a critical insight: the dramatic performance drops observed in \S~\ref{sec:fuzzing_results} (7.9\%--83.3\%) are not random occurrences, but stem from a \textit{fundamental structural flaw} in MAS architecture---the \textit{planner-coder gap}.
As presented in Table~\ref{tab:failure_distribution}, 75.3\% of failures are rooted in this gap, where semantic alignment breaks down during multi-stage information transformation from planning to coding agents. Specifically, generated plans maintain logical correctness, but coding agents fail to properly implement correct code from these plans. This gap---characterized by information loss and semantic drift across agent boundaries---explains why semantically equivalent mutations, which should not affect correctness, instead trigger systematic failures. In contrast, only 24.7\% of failures stem from other causes: 15.3\% from \textit{plan logic errors} (atomic LLM errors where planners generate algorithmically incorrect plans), and 9.3\% from \textit{invalid cases} (inherent ambiguities in original requirements). The planner-coder gap thus represents the dominant, addressable failure mode in contemporary MASs.


This finding uncovers a previously overlooked dimension that extends beyond the
current research focus. While preliminary
studies~\cite{dong2024self,zhang2024pair,hong2023metagpt} have established that
planning constitutes a critical coding step determining implementation logic and
that effective inter-agent communication is essential for correct plan
implementation, most have concentrated primarily on generating algorithmically
correct plans. Our analysis, however, demonstrates that even with logically
correct plans, the \textit{planner-coder gap} emerges as a fundamental
impediment to robust code generation in MASs. This gap manifests as a breakdown
in semantic alignment during the multi-stage information transformation process,
where the planner's high-level abstractions fail to preserve sufficient
implementation constraints for the coder to reconstruct the original intent.

Through comprehensive analysis, we identify two principal mechanisms underlying
this gap: (1) \textit{semantic drift}, where the planner's intent is
progressively diluted through abstraction, resulting in plans that are logically
sound but lack critical implementation details; and (2) \textit{context
fragmentation}, where the coder loses access to crucial constraints and boundary
conditions that were implicit in the original requirements but not explicitly
preserved in the plan. Consequently, we systematically categorize the
planner-coder gap into five distinct error patterns (EPs) and analyze their
respective distribution frequencies.

\begin{description}[leftmargin=*,noitemsep,topsep=0pt]

     \item [EP-1. Gap in Core Concepts (32.7\%).] Many failure questions involve
            specific core concepts essential to problem-solving.
            For example, the requirement of HumanEval-26\footnote{Short for
            ``the 26th question in HumanEval.'' Similar notations are used
            in the rest of the paper.} is to ``remove all duplicates in the
            input list''. Here, ``remove duplicates'' means deleting all
            elements that appear more than once. The planner provides correct
            instructions to ``remove all duplicates'' but does not clarify
            the meaning of ``duplicate''. This leaves a gap that the coder
            misinterprets ``remove duplicate'' as removing repeated
            occurrences of elements, but keeping one instance of all
            duplicates. 

    \item [EP-2. Gap in Edge Cases (19.5\%).] In some cases, the
            generated code returns incorrect results for edge cases. Although the
            plan correctly identifies edge cases using expressions such as ``handle
            the cases where...'', the coder still fails to handle them when there
            are no clear explanations of the expected results for edge cases,
            revealing the gap in edge case handling. 

    \item [EP-3. Gap in Complex Logic (15.9\%).] Coders can fail to follow
          the logic when there are insufficient concrete explanations and
          analysis. In some cases, steps in the plan contain complex logic like ``sort the coordinates of each row by columns'', which the coder fails to comprehend. 

    \item [EP-4. Gap in Relational Phrases (9.7\%).] Our case study
          demonstrates that coders are prone to misunderstand phrases that
          express quantity or spatial relationships between variables, indicating the need for
          further interpretation of these expressions. Expressions
          like ``at least as much as'', ``repeated two multiply two times''
          are often misunderstood by the coder in our experiments. 

    \item [EP-5. Gap in Condition Judgments (22.1\%).] Many
            questions require different logic paths for various situations. Code implementations sometimes omit
            certain logic paths specified in the plan when there are no detailed
            explanations of the condition for each logic path. For example,
            in HumanEval-128, the plan
            for the mutated question correctly illustrates the proper logic as
            ``calculate the product of signs and sum of all numbers''. However, the question requires special treatment when the number is zero. Without a detailed explanation for this requirement in the plan, the coder omits the logic path to check if the number is zero. 

\end{description}

\parh{Implications for Robustness Enhancement.} Each of these error patterns  highlights the structural nature of the planner-coder gap: the patterns are not attributable to prompt understanding failures at the planning stage, but rather stem from communication breakdowns between agents where the planner's high-level abstractions fail to preserve sufficient implementation details. Accounting for 75.3\% of observed failures, the planner-coder gap represents the dominant robustness issue in MASs. This finding motivates our repair approach in \S~\ref{sec:repairing}, which targets this gap through mechanisms that bridge the planner-coder interface.

\section{Augmenting MAS Robustness}
\label{sec:repairing}


Having identified the planner-coder gap as the root cause of 75.3\% of robustness failures, we now address the question: \textit{how can we mitigate
this fundamental structural flaw?} To answer this, we first formulate the
planner-coder gap by modeling the multi-stage information transformation process
in MASs. This formulation reveals that information loss accumulates across agents, providing a theoretical foundation for improvement. Based
on this formulated understanding, we propose a repairing method comprising two
components: \textit{multi-prompt generation} and \textit{monitor agent
insertion}.

\subsection{Reflection and Formulation of Planner-Coder Gap}
\label{subsection:formulation}

Our empirical discovery in \S~\ref{sec:implication} reveals that the planner-coder gap is not merely a prompt engineering issue, but a fundamental vulnerability in the multi-stage transformation architecture of MASs. To design an effective repair, we must first understand \textit{why} this gap occurs.
Unlike single LLMs that directly encode input descriptions into code, MASs employ a multi-agent workflow where the planner plays a central role in determining generation logic~\cite{zhang2024pair,hong2023metagpt,dong2024self,islam2024mapcoder}. We formulate this process to characterize where and how information loss occurs.

In MASs, the planner serves as the central orchestrator, deciding the logic of the code to generate~\cite{jiang2024self,zhang2024pair}.
As discussed in Sec.~\ref{sec:background}, the
planner functions as a critical intermediary that must bridge the conceptual gap
between human-readable requirements and coder-executable logic.
Specifically, the planner performs two critical transformations: first, it decomposes 
user inputs into a comprehensive requirement set; then, it designs the coding steps to fulfill these requirements~\cite{dong2024self,hong2023metagpt}:

\begin{align}
    \label{eq:plan1}
    \mathcal{R} &= \mathcal{A}_p^{req}(r) \\
    \label{eq:plan2}
    \mathcal{L} &= \mathcal{A}_p^{logic}(\mathcal{R})
\end{align}




\noindent where $r$ represents the input natural language requirement, $\mathcal{R} = \{r_1, r_2, ..., r_n\}$ denotes the decomposed requirement set, and $\mathcal{L} = \{l_1, l_2, ..., l_m\}$ represents the coding logic steps expressed in natural language. The generated plan includes both decomposed requirements and coding logic steps:
\begin{equation}
    p = \{\mathcal{R}, \mathcal{L}\}
\end{equation}
Instructed by the coding steps, the coder then translates each step $l_i$ into corresponding code blocks:
\begin{equation}
     \label{eq:code_generation}
    c = \mathcal{A}_c(\mathcal{L}) = \bigcup_{i=1}^{m} \mathcal{A}_c(l_i)
\end{equation}

\noindent where $c$ represents the generated code. The \textit{planner-coder gap} manifests in this transformation, where semantic misalignment occurs between the coding logic and code.
Our analysis reveals that this gap accounts for 75.3\% of failures (Table~\ref{tab:failure_distribution}), highlighting a severe robustness issue.

Fundamentally, the interaction between agents in MASs can be viewed as a cascade
of information transformations across agent boundaries, where each agent
receives, processes, and transmits information to the next stage. 
This multi-stage transformation process 
is inherently susceptible to information loss~\cite{shannon1948mathematical,cover2006elements}. 
The aforementioned issue of planner-coder gap is further exacerbated 
by the sequential nature of the transformation process, as there are multiple transformation stages (see Eq.~\ref{eq:plan1},~\ref{eq:plan2},~\ref{eq:code_generation}). 
More specifically, each transformation stage can introduce subtle semantic shifts and information loss: 
the planner might correctly identify what needs to be done but express 
it in terms that are ambiguous to the coder, or the coder might technically 
implement the logic steps but miss the underlying intent.


This issue stems from the inherent
complexity of inter-agent communication and the structural limitations of
current task distribution~\cite{li2024survey,han2024challenges}. Consequently, without robust mechanisms to promote effective communication across agents, MASs remain vulnerable to systematic
failures.

\begin{figure*}[!t]

    \centering

    \includegraphics[width=1\textwidth]{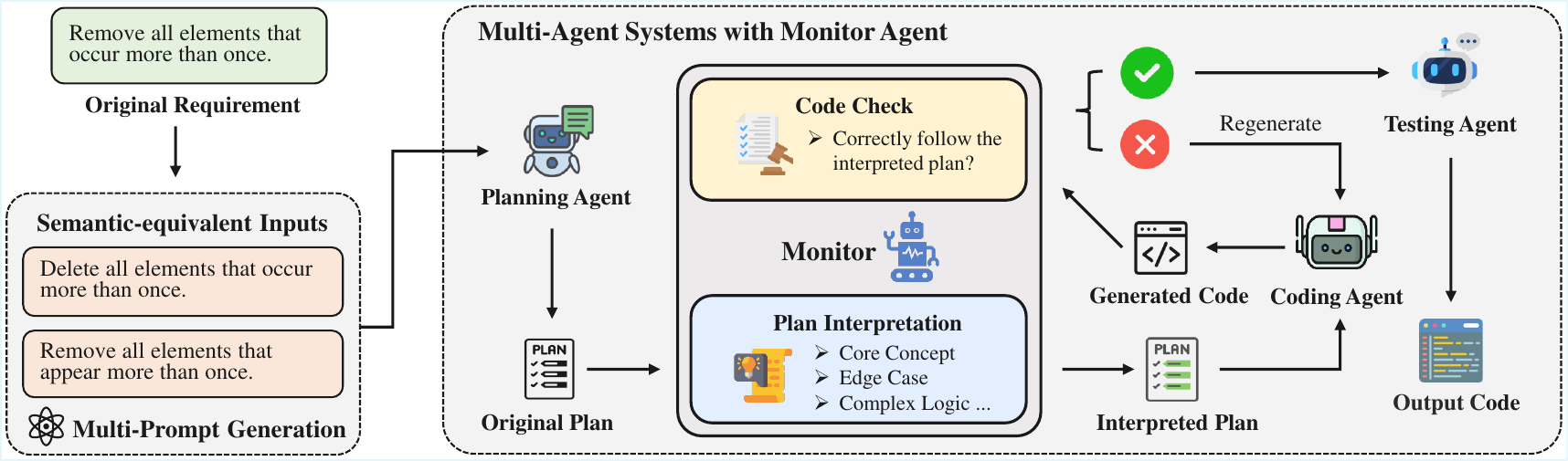}
    \vspace{-18pt}
    \caption{Overall workflow of our repairing method. \textit{Multi-prompt
            generation} generates various semantically-equivalent versions of user
        input. The monitor agent conducts \textit{plan interpretation} and
        \textit{code check} to boost communication between planner and
        coder.}
    \vspace{-10pt}
    \label{fig:repair_pipeline}

\end{figure*}

\subsection{Overview of Our Repairing Method}

To mitigate the information loss discussed in \S~\ref{subsection:formulation}, we propose a repairing method that
 enhances both the diversity of input expressions and the quality of inter-agent
 communication. Our approach comprises two principal components: \textit
 {multi-prompt generation} and \textit{monitor agent insertion}.


As illustrated in Fig.~\ref{fig:repair_pipeline}, upon receiving an input requirement, we first mutate the 
requirement and generate multiple semantic-equivalent expressions to reduce likelihood of MASs misinterpreting specific expressions. For each generation, the planner generates the 
original plan and forwards it to the monitor, a new agent we introduce to mitigate the planner-coder gap. After receiving the plan from the planner, the monitor agent 
interprets the plan by providing detailed explanations for core concepts, edge cases, complex logic, relational 
phrases, and conditional judgments (see Sec.~\ref{sec:implication}), which compensate for the information loss. Subsequently, the coder implements code based 
on this interpreted plan. In addition, the monitor also checks the implementation to enhance the alignment with the 
interpreted plan and requests regeneration when misalignments are detected.




\subsection{Multi-Prompt Generation}

Our analysis in Sec.~\ref{sec:fuzzing} demonstrated that semantic-preserving mutations can expose robustness issues in MASs. Meanwhile, as the
mutation operators explore different expressions of the questions, we observe that these mutations can also clarify ambiguous expressions, potentially improving performance on certain formulations. Leveraging this insight, we propose a multi-generation approach that explores different expressions of the same requirement by reusing
the semantic-preserving mutation operators introduced in
\T~\ref{tab:operators}.

When the MAS receives an input question, we apply
these mutation operators to generate $k$ alternative expressions as
generated prompts, with $k+1$ different versions including the original
one. During the code generation process, where $n$ represents the total
number of generation attempts, we execute the MAS with the original input
question and each mutated question for $\frac{n}{k+1}$ times.

This approach enhances robustness by leveraging diverse yet semantically
equivalent expressions, allowing the MAS to explore multiple interpretation
paths for the same requirement. In our implementation, we employ the MAS's
backend LLM for mutation generation to ensure consistency. We set $k=2$
(three total prompts including the original one) to balance
improvement gains with computational efficiency, which also demonstrates generalizability across
different MAS architectures and datasets, as shown in Sec.~\ref{sec:ablation_study}.

\begin{figure}[!tbp]

   \centering

   \includegraphics[width=0.6\textwidth]{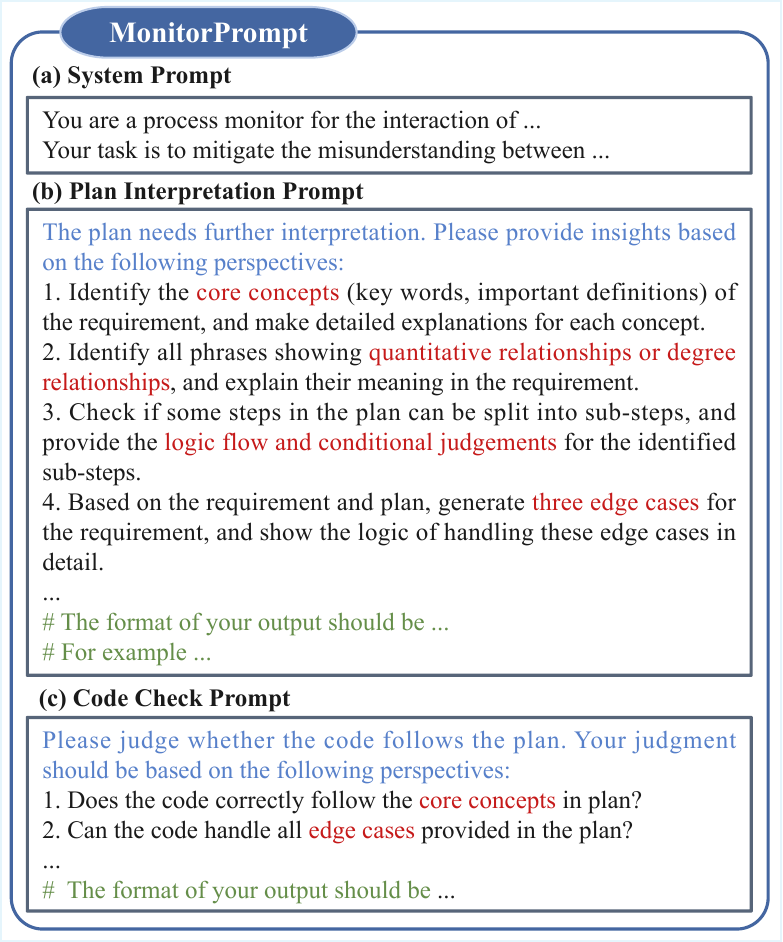}
   \vspace{-6pt}
   \caption{Prompt for the monitor agent. \textcolor{myblue}{Blue} describes the task, \textcolor{mybrown}{brown} echoes the five EPs in \S~\ref{sec:implication}, while \textcolor{mygreen}{green} provides the i/o format and examples. }
   \label{fig:prompt}
   \vspace{-18pt}

\end{figure}

\subsection{Monitor Agent Insertion}


Recent studies show that LLMs can provide evaluation highly aligned with humans~\cite{wang2025can,zheng2023judging}. Building upon this insight,
we introduce a monitor agent between planner and coder to enhance plan quality
and strengthen planner-coder alignment. This new agent performs two critical tasks: 
\textit{plan interpretation} and \textit{code check}, serving as a crucial mechanism 
to compensate for the information loss inherent in the code generation process. 
Note that this design is motivated by recent findings that specialized agents executing well-defined coordination and evaluation subtasks achieve superior reliability~\cite{fourney2024magentic,valmeekam2023planning}. By focusing on these well-defined, atomic tasks, the monitor exhibits greater 
robustness compared to the multi-stage transformations between planner and 
coder that trigger the planner-coder gap.

\parh{Plan Interpretation.}
\label{sub:plan_interpretation}
Despite remarkable efforts in previous work~\cite{dong2024self,zhang2024pair,islam2024mapcoder} to devise sophisticated prompting
structures for planning agents, they primarily focus on choosing the right algorithm without considering the information transmission loss in MASs (see~\E~\ref{eq:plan1},~\ref{eq:plan2}).
As discussed in \S~\ref{subsection:formulation}, the multi-stage transformation process of MAS introduces information loss, where the planner generates generally correct logic but discards implementation-critical details, leading to incorrect code implementations by the coder.


To effectively address this issue and enhance plan quality, we prompt the inserted
monitor agent to interpret the plan generated by the planner to make it more
comprehensible to the coder. The monitor agent aims to enhance the clarity and completeness of the planned coding logic and mitigate the planner-coder gap, rather
than correcting the semantic content of the plan. Specifically, the monitor
interprets the decomposed requirement $\mathcal{R}$ and coding logic
$\mathcal{L}$ with detailed explanations and clarifications targeting the five
error patterns identified in Sec.~\ref{sec:implication}, producing an
interpreted plan $p'$ that bridges the semantic gap between planning and coding.
Fig.~\ref{fig:prompt} presents an illustrative example of our prompt for the
monitor agent. By interpreting the plan $p$ into $p'$, the monitor facilitates the restoration of critical implementation details lost in
Eq.~\ref{eq:plan1},~\ref{eq:plan2} before code generation, which effectively
reduces the information loss between the coding logic and
final implementation. 
We employ few-shot prompting with carefully crafted examples to
guide the interpretation process, focusing on the fundamental components of
programming plans---such as core concepts, logic flows, and edge cases---rather
than being tailored to any specific problem domain. This task-agnostic design
provides general applicability across different code generation scenarios.

\sparh{Clarification.}~Note that one might consider incorporating the monitor's interpretation logic directly
into the planner's prompt. However, our proposed approach of a separate monitor
agent offers advantages in terms of modularity and scalability. Rather than
modifying the planner's core functionality, this separation of concerns allows
the monitor to be seamlessly integrated as a ``quality gate'' into various
existing MAS architectures without altering their fundamental agent logic or
task distribution. This design promotes reusability, easier maintenance, and
addresses robustness issues across different MAS implementations without
requiring architecture-specific modifications. 
We also follow the practice
 of consistent backend in a MAS to build the monitor agent with the backend LLMs of the MAS.

\parh{Code Check.}
Although the interpreted plan could preserve crucial details, the information loss between the interpreted plan $p'$ and generated code $c$ remains undetected (see \E~\ref{eq:code_generation}), posing a potential issue of misalignment.
Therefore, we
further incorporate code check as another monitoring task to check semantic alignment between the generated code and the interpreted plan.



Instead of using the testcase-based testing approaches that require dynamic execution with
 proper running environment, we directly prompt LLMs to perform static inspection and
 evaluate code compliance with the detailed specifications in $p'$,
 which has been shown to be effective with the rich contextual information provided by the interpreted plan~\cite{tong2024codejudge,zhuo2024ice}.
After the coder produces the solution code, we feed both the interpreted
plan and the code back to the monitor to check whether the code
complies with the plan regarding the different aspects of interpretation,
as shown in \F~\ref{fig:prompt}(c). If the monitor identifies mismatches
between code implementation and the plan, it returns the implementation
to the coding agent for revision. By checking alignment between the interpreted plan $p'$ and generated code $c$, the monitor creates a validation loop that detects and corrects information drift, effectively reducing the information loss that would otherwise accumulate through unchecked transformations.
For efficiency, we use
zero-shot prompting for code check and limit the process to execute
only once.

    \section{Evaluation}
\label{sec:evaluation}

We evaluate our repairing method by answering the following three research
questions:

\begin{description}[leftmargin=*,noitemsep,topsep=0pt]

    \item [RQ1:] Can our repairing method effectively mitigate failures identified in fuzzing?

    \item [RQ2:] How do different components contribute to the overall performance?

    \item [RQ3:] Can our repairing method enhance the robustness of MASs against fuzzing?


\end{description}

Specifically, RQ1 first gauges the effectiveness of our method in
repairing the failures that are identified in fuzzing. Then, we conduct an
ablation study to evaluate the individual contributions of different components
within our repairing method in RQ2. In RQ3, we re-run the fuzzing process to
evaluate whether our method enhances MASs' robustness against fuzzing.

\subsection{RQ1: Can our repairing method effectively mitigate failures
    identified in fuzzing?}

\begin{table*}
    \centering
        \centering
        \setlength{\tabcolsep}{3pt}
        \caption{RQ1: Repair results on failures identified through fuzzing.}
        \scalebox{0.7}{
            \begin{tabular}{c|c|ccc|ccc|ccc|ccc}
            \toprule
            \multirow{2}{*}{MAS}                & \multirow{2}{*}{Backend LLM} & \multicolumn{3}{c|}{{HumanEval}} & \multicolumn{3}{c|}{{MBPP}} & \multicolumn{3}{c|}{{CodeContest}}  & \multicolumn{3}{c}{{CoderEval}}                                                                                        \\
            \cmidrule(lr){3-5} \cmidrule(lr){6-8} \cmidrule(lr){9-11} \cmidrule(lr){12-14}
                                           &                              & Total                               & Solved                         & Rate                             & Total & Solved & Rate  & Total & Solved & Rate  & Total & Solved & Rate \\
                \midrule
                \multirow{3}{*}{SCCG}      & GPT-3.5                      & 39                                  & 30                             & 76.9\%                            & 60    & 48     & 80.0\% & 12    & 9      & 75.0\% &  11   &  9    & 81.8\%\\
                                           & GPT-4o                       & 22                                  & 11                             & 50.0\%                            & 49    & 21     & 42.9\% & 21    & 13     & 61.9\% &   12  &  5    & 41.7\%\\
                                           & Deepseek                     & 28                                  & 15                             & 53.6\%                            & 63    & 34     & 53.9\% & 12    & 8      & 66.7\% &    15 &   7   & 46.7\%\\
                \midrule
                \multirow{3}{*}{MetaGPT}   & GPT-3.5                      & 49                                  & 37                             & 75.5\%                            & 38    & 23     & 60.5\% & 9     & 5      & 55.6\% &  10   &  8     & 80.0\%\\
                                           & GPT-4o                       & 24                                  & 13                             & 54.2\%                            & 37    & 16     & 43.2\% & 20    & 13     & 65.0\% &  9   & 5     & 55.6\%\\
                                           & Deepseek                     & 24                                  & 17                             & 70.8\%                            & 38    & 20     & 53.3\% & 21    & 14     & 66.7\% &  9   &  8    & 88.9\%\\
                \midrule
                \multirow{3}{*}{PairCoder} & GPT-3.5                      & 15                                  & 8                              & 53.3\%                            & 25    & 14     & 56.0\% & 22    & 9      & 40.9\% &   19  &  10   & 52.6\%\\
                                           & GPT-4o                       & 14                                  & 7                              & 50.0\%                            & 25    & 16     & 64.0\% & 16    & 7      & 43.8\% &    20 &   14   & 70.0\%\\
                                           & Deepseek                     & 14                                  & 7                              & 50.0\%                            & 26    & 14     & 53.8\% & 20    & 8      & 40.0\% &    12 &   9   & 75.0\%\\
                \bottomrule
            \end{tabular}
        }
        \label{tab:RQ1_repair_rate}
    \vspace{-5pt}
\end{table*}

To address RQ1, we evaluate the performance of our repairing method on the
failed questions identified during fuzzing. We collect the failures for each 
MAS with different backend LLMs to form our evaluation dataset. Since all 
MASs initially failed to solve these questions, the baseline success rate is 
zero. After applying our repairing method, we re-evaluate the MASs on these questions.


Table~\ref{tab:RQ1_repair_rate} presents the compelling repair results: our method 
enables MASs to successfully solve 40.0\%--88.9\% of the questions they 
previously failed. This significant improvement demonstrates that enhanced 
expression diversity and better planner-coder communication effectively reduce 
information loss during agent communication and enhance the robustness of MASs.
Taking HumanEval-18 as an example, the original plan of MetaGPT (Deepseek) only
mentions ``please handle the edge case when the substring is empty,''
leading to the coder's misunderstanding. In contrast, the interpretation
generated by the monitor further explains the logic of this case by
providing examples and a detailed logic flow. Consequently, the coder
successfully implements the correct code.

\begin{table}
    \caption{Repair ratios for different categories of failures from SCCG (GPT-3.5).}
    \centering
    \scalebox{0.8}{
        \begin{tabular}{c|ccccc|c|cc}
            \toprule
            Category & EP-1 & EP-2 & EP-3 & EP-4 & EP-5 & Overall & Plan Logic Errors & Invalid \\
            \midrule
            Solved (\%) & 80.0 & 83.3 & 80.0 & 100.0 & 85.7 & 83.9 & 57.1 & 0.0 \\
            \bottomrule
        \end{tabular}
    }
    \label{tab:RQ1_repair_rate_category}
    \vspace{-13pt}
\end{table}


The improvement contributed by our method varies across different MAS architectures. As shown in \T~\ref{tab:RQ1_repair_rate},
SCCG and MetaGPT show exceptional improvement, achieving repair rates 
exceeding 80\%. PairCoder shows more modest gains of 40\%--75\%.  
Given that PairCoder incorporates plan selection strategies, the readability and quality of plans generated by
PairCoder surpass those of other MASs, therefore suffering less from the planner-coder gap.
Similarly, the choice of backend LLM influences repair effectiveness. MASs 
using GPT-3.5 or Deepseek as backend LLM benefit more from our method compared to those 
using GPT-4o. These results
indicate that our method is more beneficial for less capable MASs
with less effective backend LLMs, which typically struggle more with plan comprehension and are more
susceptible to the planner-coder gap.




To further understand the strengths of our method,
\T~\ref{tab:RQ1_repair_rate_category} presents the repair
results for different EPs identified in SCCG (GPT-3.5). 
Our repairing method demonstrates remarkable effectiveness in
handling failures originating from the planner-coder gap, successfully 
resolving 83.9\% of such cases. 
Notably, the consistently high repair ratios across all 
five EPs (80.0\%--100.0\%) indicate that our repairing method serves as a generalizable 
strategy for bridging the planner-coder gap, regardless of the 
specific type of error pattern involved.
Surprisingly, our method also demonstrates considerable effectiveness 
on plan logic errors, successfully repairing 57.1\% of such cases. 
Although our approach primarily targets planner-coder gap, this result suggests that the interpretation process may also compensate
for the planner's misunderstandings, as explicit interpretation may inadvertently 
correct or clarify flawed reasoning in the original plan.

\begin{tcolorbox}[ size = small ]
    After adopting our method, MASs successfully solve 40.0\%--88.9\% of
    failures identified during fuzzing. Our repairing method demonstrates remarkable
 effectiveness in handling failures originating from the planner-coder gap, 
 successfully repairing the five EPs identified in \S~\ref{sec:implication}.
\end{tcolorbox}


\subsection{RQ2: How do different components contribute to the overall performance?}
\label{sec:ablation_study}

\begin{figure*}[t]
    \centering
    \includegraphics[width=1\textwidth]{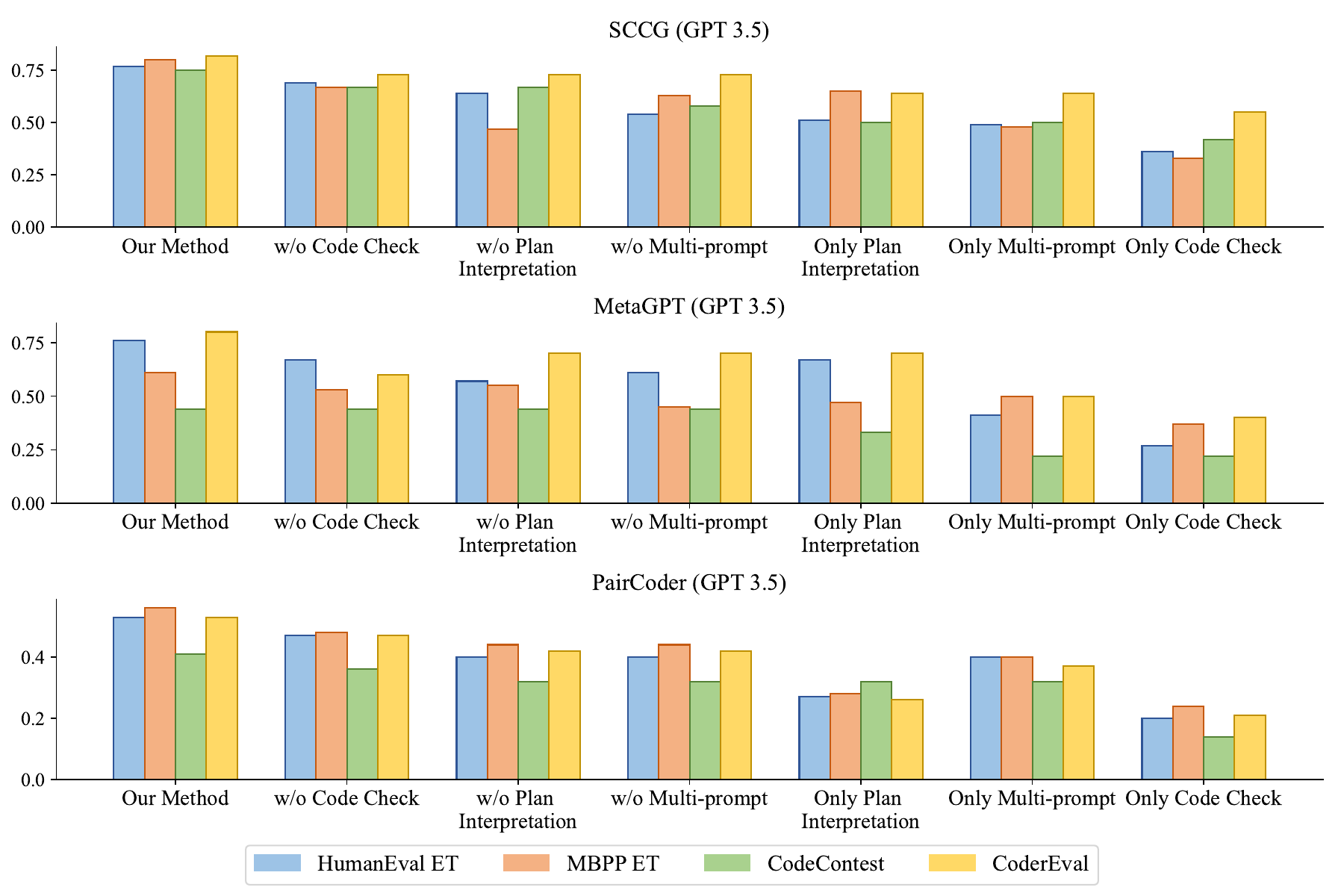}
    \vspace{-23pt}
    \caption{Comparison of repairing performance when removing different
        components.}
    \vspace{-22pt}
    \label{fig:RQ2_combined}
\end{figure*}

\begin{figure*}[!h]
    \centering
    \includegraphics[width=1\textwidth]{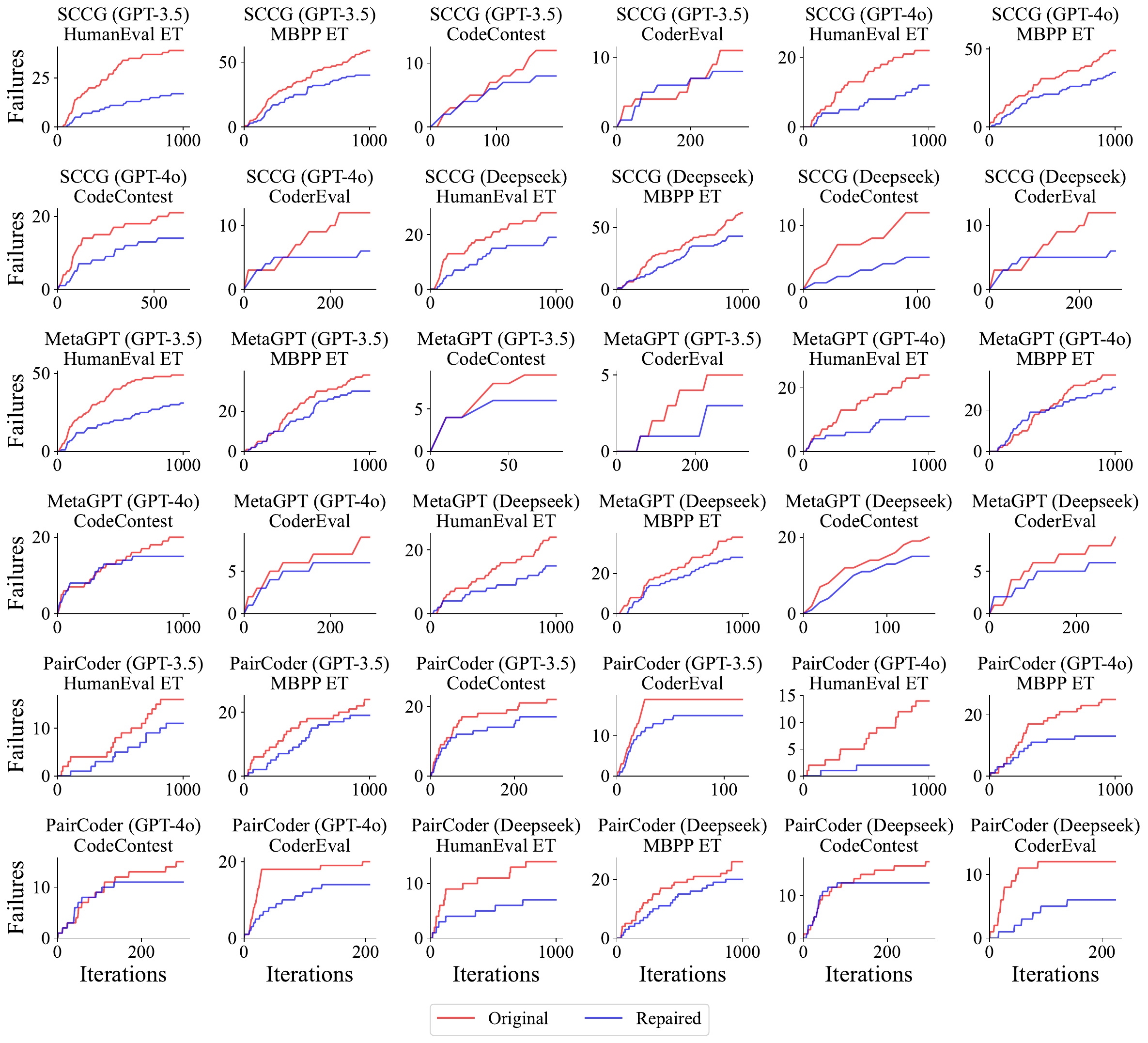}
    \vspace{-20pt}
    \caption{Comparison of failures of the original and repaired MASs found in
        fuzzing.}
    \vspace{-16pt}
    \label{fig:RQ3}
\end{figure*}

We conduct ablation studies to investigate the effectiveness of multi-prompt
 generation and monitor agent insertion in our repairing method. For monitor
 agent insertion, we separately examine ``Plan Interpretation'' and ``Code
 Check'' to provide in-depth analysis. We present results with GPT-3.5 as the
 backend LLM, as it represents a typical configuration with other LLMs
 exhibiting similar patterns. 

\F~\ref{fig:RQ2_combined} shows that removing any component degrades repair
performance across all three MASs, indicating that all components contribute
meaningfully and our design outperforms all alternative configurations. While disabling ``Code Check'' causes the smallest performance
decrease, enabling ``Code Check'' still contributes to the overall repair
performance and ``Only Code Check'' contributes approximately 40\% of our
method's full repair capability, demonstrating that post-generation validation
effectively catches misalignments even without enhanced plan interpretation.

Moreover, our results reveal distinct patterns across MAS
architectures, where the
relative importance of ``Plan Interpretation'' versus ``Multi-prompt'' varies by
MAS implementations. In MetaGPT, ``Only Plan Interpretation'' substantially
outperforms ``Only Multi-prompt'' because MetaGPT generates abstract plans that
challenge coder comprehension, where interpretation bridges critical
understanding gaps. Conversely, PairCoder benefits more from ``Only
Multi-prompt'' since its plans have better readability, making the diversity of
input expressions from multi-prompt generation more beneficial. These findings
demonstrate that both strategies are valuable as different MASs exhibit unique
flaws: MASs like MetaGPT with abstract plans benefit most from interpretation,
while those like PairCoder with clear but rigidly expressed plans gains more
from prompt diversity.

\begin{tcolorbox}[
        size = small
    ]
    Our repairing method achieves optimal performance among all ablation configurations, with each component 
    mitigating different information loss and enhancing the overall repair capability.
\end{tcolorbox}

\subsection{RQ3: Can our repairing method enhance the robustness of MASs against fuzzing?}


While RQ1 demonstrates that our repairing method can effectively fix failures identified during the initial fuzzing process, it remains unclear whether our repairing method can prevent or reduce new failures when subjected to continued fuzzing.
Therefore, in this RQ, we apply our repairing method to all MASs and re-execute the
fuzzing process proposed in Sec.~\ref{sec:fuzzing}.
\F~\ref{fig:RQ3} shows the number of failures discovered from the original MASs and the repaired ones through fuzzing.


From \F~\ref{fig:RQ3}, we observe that on all three MASs, the speed of
discovering failures (i.e., the slope of the blue line and the red line) slows
down, and identified failures decrease after applying our repairing method, showing
that the repaired MASs exhibit superior robustness compared to their original
counterparts.
Notably, PairCoder
(GPT-4o) exhibits the most remarkable robustness improvement against
fuzzing, with 85.7\% of failures in HumanEval eliminated.
SCCG shows stable results across different backend LLMs and datasets, with
28.6\%--53.8\% decrease in failure numbers. These results indicate that our
repairing method can effectively enhance the robustness of MASs against
semantic-preserving mutations, resulting in fewer failures discovered
during fuzzing.



Across different datasets, MASs demonstrate the most significant robustness
improvement on HumanEval, with failures decreasing by 31.2\%--85.7\%,
surpassing other datasets. This superior performance can be
attributed to HumanEval's well-defined questions with clear logic
structures. Therefore, our repairing method effectively enhances
the coder's comprehension of the tasks by accurately capturing the
question's intent, thereby substantially improving MAS robustness.


\begin{tcolorbox}[size=small]
 Our repairing method enhances MAS robustness
 against fuzzing, with up to 85.7\% failure number reductions.
 Repaired MASs exhibit superior robustness on the HumanEval dataset.
\end{tcolorbox}

\section{Discussion}


\parh{Time Cost.}~Time efficiency is a critical factor in MASs, 
as computational overhead directly impacts system scalability and practical deployment~\cite{wu2023survey}.
To understand the time cost of our repairing method, we
measure the average execution time per attempt with and without our repairing method.
Table~\ref{tab:time_analysis} presents the experiment results on representative setting (HumanEval, GPT-3.5), as other settings (e.g., datasets, LLMs) show similar trends.

\begin{table}[t]
 \centering 
 \caption{Comparison of time cost on HumanEval with GPT 3.5 as backend LLM.}
 \label{tab:time_analysis}
 \scalebox{0.8}{
 \begin{tabular}{ccccc}
 \toprule
  & Our method & w/o Multi-prompt & w/o Monitor & Original \\
 \midrule
 SCCG & 13.6s & 12.9s & 11.3s & 10.9s \\
 MetaGPT & 18.4s & 17.6s & 15.3s & 14.8s \\
 PairCoder & 20.1s & 19.0s & 17.3s & 16.4s \\
 \bottomrule
 \end{tabular}
 }
 \vspace{-15pt}
\end{table}


In general, the additional time overhead primarily comes from the
two components of our repairing method: multi-prompt generation and monitor
agent insertion. For multi-prompt generation, since we maintain the same
total number of generation attempts $n$, the primary overhead comes only
from generating $k$ mutated prompts, which incurs little additional
overhead for API calls compared to the main generation process. Besides,
the monitor agent introduces two additional API calls per generation
attempt, taking merely 2.7s--3.7s across different MASs.
We clarify that this overhead is modest in real-world application
scenarios, given that a full MAS execution typically takes more than 30 seconds. On the other hand, the adoption of the monitor
agent brings substantial robustness improvements, as shown in
\S~\ref{sec:evaluation}, making the cost-benefit trade-off favorable. 







\parh{Threats to Validity.} We identified and mitigated several potential threats to validity in our evaluation.
The first threat is the representativeness of our experimental setup. To 
address this, we evaluated three popular MASs~\cite{dong2024self,zhang2024pair,
hong2023metagpt} with three distinct LLMs~\cite{openai2023chatgpt,achiam2023gpt,
zhu2024deepseek} from different model families across four diverse datasets~\cite
{chen2021evaluating,austin2021program,li2022competition,yu2024codereval}. 
The second threat is the reliability of semantic-preserving mutation. 
To mitigate this threat, we conducted manual verification confirming that 99.2\% of mutants preserve semantics. Additionally, we measure the semantic similarity using Sentence-Transformers~\cite{reimers2019sentence}, whose results show 98.4\% average similarity.
Thus, the robustness issues revealed by our fuzzing are meaningful. The last threat is subjectivity in manual failure
analysis. To ensure reliable implications from failure cases, two software
developers with over five years of programming experience independently categorize
failure reasons, cross-check the annotation, and resolve all disagreements.
    \section{Related Work}

\parh{MASs for Code Generation.}~
Researchers have designed various MASs with diverse architectures by defining 
different roles among multiple LLM agents~\cite{hong2023metagpt,dong2024self,zhang2024pair,islam2024mapcoder,islam2025codesim,jiang2024self}.
For instance, MetaGPT~\cite{hong2023metagpt} employs five distinct agents to 
simulate a software company's development workflow, while the Self-Collaboration~\cite{dong2024self} uses three agents for planning, coding, and testing.
To enhance plan correctness, 
Zhang et al.~\cite{zhang2024pair} use clustering to select the optimal plan, 
and Islam et al.~\cite{islam2024mapcoder} employ dynamic traversal to assign confidence 
scores to plans.
However, these methods focus primarily on logical correctness while neglecting the
communication gap between planner and coder, which is first identified in
this paper. In this work, we propose a repairing method to address the robustness issue of MASs by mitigating the gap.

\parh{Testing Code Generation Models.}~Numerous studies have applied testing
methodologies to evaluate code generation LLMs. Wang et
al.~\cite{wang2023recode} introduce ReCode, a benchmark for robustness
evaluation through systematic perturbation testing, while Mastropaolo et
al.~\cite{mastropaolo2023robustness} conduct empirical studies on GitHub
Copilot's robustness in real-world scenarios.
Recent advances have proposed diverse methods to evaluate
code generation models under different scenarios~\cite{chen2024nlperturbator,yang2025assessing,zhong2024can,chen2023evaluating}, like adversarial
training~\cite{zhang2024codefort} and instruction
concretization~\cite{yan2025robustness}.
However, none of these works focus on MASs and their robustness remains
under-explored. We conducted the testing-based study over MAS robustness and
identified the planner-coder gap as the main cause of robustness issues.

    \section{Conclusion}
This paper presents the first comprehensive study on the robustness of MASs for code generation, revealing their robustness issue on semantically equivalent mutations. We identify the ``planner-coder gap'' as the root cause and propose a novel repair method incorporating multi-prompt generation and monitor agent insertion. Our findings provide essential insights for developing more reliable MASs and establish a foundation for future robustness research in this domain.

    \bibliographystyle{ACM-Reference-Format}
    \bibliography{references}
\end{document}